\documentclass[preprint,floatfix,showpacs,preprintnumbers,amsmath,showkeys,amssymb,aps,pra]{revtex4}

\usepackage{graphicx}
\usepackage{dcolumn}
\usepackage{bm}

\begin{document}

\title{Nuclear-resonant electron scattering}

\author{Adriana P\'alffy}
\email{Palffy@mpi-hd.mpg.de}

\author{Zolt\'an Harman}
\email{Harman@mpi-hd.mpg.de}
\affiliation{Max-Planck-Institut f\"ur Kernphysik, Saupfercheckweg 1,
69117 Heidelberg, Germany}

\date{\today}

\begin{abstract}

We investigate nuclear-resonant electron scattering as occurring in the 
two-step process of nuclear excitation by electron capture (NEEC) 
followed by internal conversion. The nuclear excitation and decay are 
treated by a phenomenological collective model in which nuclear states 
and transition probabilities are described by experimental parameters. 
We present capture rates and resonant strengths for a number of heavy 
ion collision systems considering various scenarios for the resonant 
electron scattering process. The results show that for certain cases 
resonant electron scattering can have significantly larger resonance 
strengths than NEEC followed by the radiative decay of the nucleus. We 
discuss the impact of our findings on the possible experimental 
observation of NEEC.

\end{abstract}

\pacs{34.80.Lx, 34.80.Dp, 23.20.Nx, 23.20.-g}

\keywords{electron scattering, nuclear excitation, resonant transitions, 
highly charged ions}

\maketitle

%%%%%%%%%%%%%%%%%%%%%%%%%%%%%%%%%%%%%%%%%%%%%%%%%%%%%%%%%%%%%%%%%%%%%%%%

\section{Introduction}

%%%%%%%%%%%%%%%%%%%%%%%%%%%%%%%%%%%%%%%%%%%%%%%%%%%%%%%%%%%%%%%%%%%%%%%%

Electron scattering on ions and atoms provides a valuable experimental 
means in both atomic and nuclear physics, and is also of great interest 
due to the significance of this process in high-temperature plasmas 
(see, for instance, Ref.~\cite{Williams} and references therein). The 
elastic scattering of free and quasifree electrons on energetic ions has 
been studied both theoretically \cite{Bartschat,Bhalla,Johnson,Shingal}) 
and experimentally 
\cite{Zouros1,Zouros2,Grabbe,Hagmann,Greenwood,Huber_spectrometer}. A 
number of studies have been carried out on electron impact excitation 
\cite{Huber,Janzen,Jiang,Suzuki,Wallbank} and resonant electron 
scattering \cite{Itoh,Badnell,Kollmar}. The latter corresponds to 
dielectronic capture, i.e., continuum electron capture by the excitation 
of a bound electron, followed by the Auger decay of the autoionizing 
state. Since the Auger rates are involved in both steps of this process, 
the corresponding cross sections are particularly sensitive to the 
electron-electron interaction. This feature can be used for studies of 
the relativistic interaction of electrons in the strongest binding 
nuclear fields available up to now. As an example, the relative 
contribution of the Breit current-current interaction to the cross 
section of resonant excitation on hydrogen-like uranium ions was shown to 
be approximately twice as large as in the case of dielectronic capture 
followed by radiative de-excitation \cite{Kollmar}.

Elastic electron scattering provides an indispensable tool for surveying 
the electromagnetic structure of ground and excited states of nuclei. 
Electron scattering as a nuclear probe has the major advantage that the 
interaction is electromagnetic and hence well known. As a result, for a 
specific charge distribution, the elastic electron scattering cross 
section can be calculated by phase-shift analysis techniques 
\cite{deVries1,deVries2}. As it will be argued in this paper, nuclear-resonant 
electron scattering in highly charged ions can even 
provide information about nuclear transitions and excited 
states via the process of nuclear excitation by electron capture (NEEC).

\begin{figure}[Ht]
\begin{center}
\includegraphics[width=0.45\textwidth]{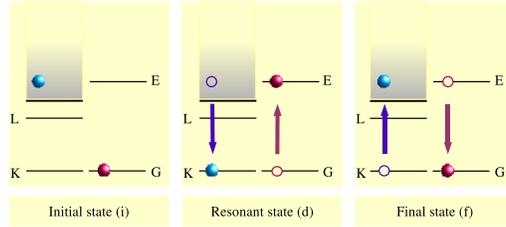}
\caption{\label{scheme1}(Color online) NEEC recombination mechanism of a 
continuum electron into the $K$ shell of an initially bare ion, followed 
by IC of the excited nucleus. The nuclear transition from the ground 
state $G$ to the first excited state $E$ is pictured schematically on 
the right-hand side of each panel.}
\end{center}
\end{figure} 

In the resonant process of NEEC, the collision of a highly charged ion 
with a free electron with matching kinetic energy leads to a resonant 
capture into an atomic orbital with the simultaneous excitation of the 
nucleus \cite{Palffy1}. This recombination process was first 
theoretically proposed by the authors of Ref.~\cite{Goldanskii} in the 
context of laser-produced plasmas, and is the time-reversed process of 
internal conversion (IC). Although not yet experimentally observed, NEEC 
has been an interesting subject after experimental observations of 
atomic physics processes with regard to the structure of the nucleus 
have been recently reported, such as bound-state internal 
conversion~\cite{Carreyre} and its time-reversed process of
nuclear excitation by electron transition~\cite{Kishimoto}, or the so-called
electronic-bridge process, which can be regarded as bound-state internal 
conversion accompanied by photon emission~\cite{Kekez85,Kalman91,Kalman01}.
Se\-ve\-ral theoretical studies have been made 
concerning NEEC in plasmas \cite{Goldanskii,Harston} or in solid targets 
\cite{Cue,Kimball1,Kimball2}.

As the electron capture in NEEC results in the excitation of the 
nucleus, $\gamma$ decay of the nucleus or IC are expected in the second 
step of the process. Several theoretical aspects of NEEC followed by the 
radiative decay of the excited nucleus have been recently addressed 
\cite{Palffy1,Palffy2,Palffy3,lifetime_prolongation}, providing 
theoretical cross sections and discussing the possible experimental 
observation of NEEC by detecting the photons emitted in the nuclear 
$\gamma$ decay. In this paper we would like to draw the attention to a 
two-step process in which NEEC is followed by IC, resulting in  
nuclear-resonant electron scattering (NRES), as schematically pictured 
in Figure~\ref{scheme1}.

Our motivation in investigating this electron scattering mechanism is 
twofold. First, nuclear-resonant electron scattering is far more 
sensitive to the electron-nucleus interaction than NEEC followed by the 
radiative decay of the nucleus, due to the presence of the IC rate in 
each of the two steps of the process. This makes NRES a more suitable 
candidate for exploring the spectral properties and dynamics of heavy 
nuclei by the use of experimental methods and facilities primarily 
developed for atomic physics. Especially, NRES may allow the 
determination of nuclear transition energies and transition 
probabilities, the study of atomic vacancy effects on nuclear lifetime 
\cite{lifetime_prolongation} and population mechanisms of excited 
nuclear levels. A second aspect concerns the experimental observation of 
NEEC.  Theoretical calculations for NEEC followed by IC or radiative 
decay of the nucleus occurring in scattering measurements are 
particularly useful in finding candidate isotopes and transitions 
suitable for experimental observation. For a number of heavy nuclei, the 
IC rates for low-lying first excited levels are substantially higher than the
radiative decay rates, with the immediate consequence that NRES cross 
sections and resonance strengths are larger than the corresponding 
values for NEEC followed by the radiative decay of the nucleus. 
Furthermore, in storage ring experiments aiming at the observation of NEEC by detecting the recombined ions 
 (as in the case of, e.g., dielectronic recombination 
experiments~\cite{Bra02,Bra03}) both nuclear decay channels should be 
taken into account. The interest for electron scattering and 
recombination experiments at the present and future storage ring 
facilities of the GSI Darmstadt \cite{GSIdesign_rep,NIMB} makes nuclear-resonant 
electron scattering in heavy highly charged ions an important 
issue for the experimental observation of NEEC.

\begin{figure}
\begin{center}
\includegraphics[width=0.45\textwidth]{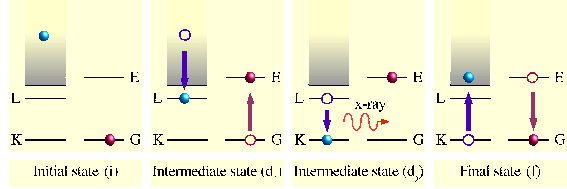}
\caption{\label{scheme2} (Color online) NEEC recombination mechanism of a continuum 
electron into the $L$ shell of an initially bare ion, followed by fast 
x-ray emission from the electronic decay to the $K$-shell ground state. 
The process is completed by IC of the excited nucleus. See text for 
further explanations.}
\end{center}
\end{figure}

In this paper we theoretically investigate resonant scattering of 
electrons undergoing NEEC followed by IC in two possible 
cases. In the first considered scenario, NEEC and IC occur in the same 
atomic orbital, as presented schematically in Figure~\ref{scheme1}. The 
continuum electron in the initial and final state has then the same 
kinetic energy in the center-of-mass reference frame. A second 
possibility considers the case in which the electron is captured into an 
excited state. NEEC then leads to a doubly-excited intermediate state 
$d_1$, as depicted in Figure~\ref{scheme2}. This intermediate state can 
decay via emission of photons from the electron shell or the nucleus, or, 
alternatively, via IC. Since x-ray emission associated with the 
electronic de-excitation is faster than nuclear decay of the low-lying 
excited nuclear states considered here, a second intermediate state 
$d_2$ in which the electron is in the ground state is reached. This 
process was denoted as nuclear excitation by electron capture followed 
by fast x-ray decay (NEECX) in an earlier 
work~\cite{lifetime_prolongation}, in analogy to the already established 
notation for the atomic process of resonant transfer and excitation 
followed by x-ray emission (RTEX). Following NEECX, in the nuclear decay 
step IC of the excited nucleus occurs if energetically allowed, 
resulting in a final state characterized by a continuum electron with 
different kinetic energy than the one in the initial state. We denote 
the process of NRES with fast x-ray decay of the captured electron by 
NRESX, in analogy to NEECX and RTEX. This more complicated three-step 
process is considered because of the advantages for the experiment 
observation due to the much broader width of the state the electron is
captured into, as it will be discussed in detail in Section~\ref{results}.

We present total cross sections and resonance strengths for NRES and 
compare them with the ones presented in Ref.~\cite{Palffy1,Palffy2} for 
the case of NEEC followed by the $\gamma$ decay of the nucleus.  The 
total cross section derivation and a brief description of the 
electron-nucleus interaction matrix elements are given in Section II. The 
electric and magnetic electron-nucleus interactions are considered 
explicitly and the nucleus is described by the help of a nuclear 
collective model \cite{Greiner}. The dynamics of the electrons is 
governed by the Dirac equation as required in the case of high-$Z$ 
elements. Section III presents the numerical results for NRES cross 
sections and resonance strengths and discusses issues of the possible 
experimental observation of NEEC. We conclude with a short summary. 
Atomic units are used throughout this paper unless otherwise specified.

%%%%%%%%%%%%%%%%%%%%%%%%%%%%%%%%%%%%%%%%%%%%%%%%%%%%%%%%%%%%%%%%%%%%%%%%%%%%%%

\section{\label{theory} Theoretical formalism }

%%%%%%%%%%%%%%%%%%%%%%%%%%%%%%%%%%%%%%%%%%%%%%%%%%%%%%%%%%%%%%%%%%%%%%%%%%%%%%

In this section we present the total cross section for the process 
of NEEC followed by IC of the excited nucleus, derived by means of 
a perturbative expansion of the transition operator. We consider the 
nuclear transition from the ground state to the first excited state with 
the simultaneous capture of a free electron into a bare ion or an ion 
with a closed-subshell configuration.  For the cases in which the 
electron capture does not occur in the ground state, the subsequent fast 
electronic x-ray decay is taken into account.

\subsection{Total cross section}
%%%%%%%%%%%%%%%%%%%%%%%%%%%%%%%%%%

The total cross section for NRES can be written with the help of the 
perturbation expansion of the transition operator, following our 
formalism presented in Ref.~\cite{Palffy1}. In order to clearly identify 
the terms contributing to the process under study, in ~\cite{Palffy1} we 
introduced Feshbach projection operators that separate the Fock space 
into subspaces corresponding to the possible initial, intermediate and 
final states. The initial state of the ion-electron system consisting of 
the nucleus in its ground state, the free electron, and the vacuum state 
of the electromagnetic field,  can be written as a direct product of the 
state vectors,
\begin{equation}\label{psi_i}
| \Psi_i \rangle = | N_i I_i M_i , \vec{p}_i m_{s_i}^+,0  \rangle \equiv
| N_i I_i M_i\rangle \otimes | \vec{p}_i m_{s_i}^+ \rangle \otimes | 0 \rangle \,.
\end{equation}
Here, $\vec{p}_i$ is the asymptotic initial momentum of the electron, $m_{s_i}$ 
its spin projection, and $N$ denotes the nuclear ground state, 
characterized by the angular momentum $I_i$ and the magnetic quantum 
number $M_i$.

For the two-step process of NEEC followed by IC of the 
excited nucleus, where the electron capture occurs into the electronic 
ground state, the intermediate state $|\Psi_d\rangle$ is given by
\begin{equation}\label{eq:dstate}
| \Psi_d \rangle = | N_d I_d M_d, n_d\kappa_d m_d,0  \rangle \equiv 
| N_d I_d M_d \rangle \otimes | n_d\kappa_d m_d \rangle \otimes | 0 \rangle \,,
\end{equation}
with $n_d$, $\kappa_d$, and $m_d$ being the principal quantum number, 
Dirac angular momentum, and magnetic quantum number of the bound 
one-electron state, respectively. The one-electron state is written in the
spherical bispinor form
\begin{equation}
\langle \vec{r}| n_d\kappa_d m_d  \rangle = \psi_{n_d\kappa_dm_d}
(\vec{r})=\left(\begin{array}{c} g_{n_d \kappa_d}(r)\Omega_{\kappa_d}^{m_d}
(\theta,\varphi)\\ i f_{n_d \kappa_d}(r)\Omega_{-\kappa_d}^{m_d}(\theta,\varphi)\end{array}
\right)\ ,\label{bound}
\end{equation}
where the $\Omega_{\kappa_d}^{m_d}$ are the spherical spinors 
\cite{Eichler} and $\theta$ and $\varphi$ are the polar and azimuthal 
angles associated with the vector $\vec{r}$, respectively. The excited 
nuclear state is denoted by $|N_d I_d M_d\rangle$.

The final state for the two-step resonant 
electron scattering on nuclei is then characterized by the nucleus in 
its ground state and the electron in the continuum,
\begin{equation}
| \Psi_f \rangle = | N_f I_f M_f, \vec{p}_f m_{s_f}^-,0 \rangle \equiv 
| N_f I_f M_f \rangle \otimes | \vec{p}_f m_{s_f}^- \rangle \otimes | 0 \rangle \, .
\end{equation}

We denote the energy eigenvalues of the states introduced above as $E_i$, $E_d$ and
$E_f$, respectively. Furthermore, the initial and final state continuum electronic wave 
functions are given in the coordinate space representation as
the partial wave expansion~\cite{Eichler}
\begin{equation}\label{partialwave}
|\vec{p} m_s^\pm \rangle=\sum_{\kappa m m_l}i^l e^{\pm i\Delta_{\kappa}}
Y_{l m_l}^*(\Omega_p)
C\left(l\ \frac{1}{2}\ j;m_l\ m_s \ m\right)| E_c \kappa m\rangle\ ,
\end{equation}
where $E_c$ is the energy of the continuum electron measured 
from the ionization threshold, $E_c=\sqrt{p^2c^2+c^4}-c^2$. The 
orbital angular momentum of the partial wave is denoted by $l$ and the 
corresponding magnetic quantum number by $m_l$. The $+(-)$ sign of the 
partial wave phases $\Delta_{\kappa}$ corresponds to the initial (final) 
free electron and the phases are chosen so that the continuum wave 
function fulfills the boundary conditions of an incoming (outgoing) 
plane wave and an outgoing (incoming) spherical wave. The total angular 
momentum quantum number of the partial wave is $j=|\kappa|-\frac{1}{2}$ 
with its projection $m=m_l+m_s$ and the symbol $C\left(j_1,j_2,j_3;m_1,m_2,m_3\right)$
stands for the Clebsch-Gordan coefficient. The partial wave functions are 
represented as
\begin{equation}
\langle \vec{r}| E_c \kappa m \rangle =
\psi_{E_c \kappa m}(\vec{r})=
\left(\begin{array}{c} g_{E_c \kappa}(r)\Omega_{\kappa}^{m}(\theta,\varphi)\\
i f_{E_c \kappa}(r)\Omega_{-\kappa}^{m}(\theta,\varphi)\end{array}
\right)\ .\label{cont}
\end{equation}

Following the formalism presented in our previous work~\cite{Palffy1}, 
the total cross section as a function of the total initial energy  
$E$ for the two-step process $i\to d\to f$ is given by
\begin{equation}
\sigma_{i \to d \to f}(E) = \frac{2\pi^2}{p^2}
\frac{A_{\rm IC}^{d \to f} Y_n^{i \to d}}{\Gamma_d} L_d(E-E_d) \, ,
\label{tcs1}
\end{equation}
where $A_{\rm IC}^{d \to f}$ is the IC decay rate of the nuclear excited 
state and $Y_n^{i \to d}$ the NEEC rate. In the denominator, $\Gamma_d$ 
denotes the total natural line width of the nuclear excited state, given 
by the sum of the partial IC and $\gamma$ decay widths, 
$\Gamma_d=\Gamma_{\rm IC}+ \Gamma_{\gamma}$. The continuum electron 
energy dependence is given by the well-known Lorentz line profile 
function
\begin{equation}
L_d(E_c - E_{\rm exc} - \varepsilon_{n_d \kappa_d}) = L_d(E-E_d) = 
\frac{\Gamma_d / 2\pi}{(E-E_d)^2 + \frac{1}{4} \Gamma_d^2}
\end{equation}
%%
%centered on the resonance energy $E_d$ and 
with the width $\Gamma_d$ given by the natural width of the excited 
nuclear state. Here we introduced the notation $E_{\rm exc}$ for the 
nuclear excitation energy and $\varepsilon_{n_d \kappa_d}$ for the 
energy of the bound intermediate electronic state. The cross section 
formula (\ref{tcs1}) is valid in the resonant case, i.e. for continuum 
electron energies $E_c$ approximately fulfilling the resonance condition 
$E_c = E_{\rm exc}+\varepsilon_{n_d \kappa_d}$.

Since NEEC is 
the time-reversed process of IC, and we consider cases involving  
transitions  between two nuclear levels only, the rates of the two processes 
occurring between the states $d$ and $f$ and $i$ and $d$ can be related 
by the principle of detailed balance,
\begin{equation}
A^{d \to f}_{\rm IC}=\frac{2(2I_i+1)}{(2I_d+1)(2j_d+1)}Y_n^{i\to d} \, ,
\end{equation}
where $j_d$ denotes the total angular momentum of the bound electron.

For the resonant three-step process depicted in Figure~\ref{scheme2}, in 
which NEEC occurs into an excited electronic state with subsequent fast 
x-ray emission, the first intermediate state given in 
Eq.~(\ref{eq:dstate}) becomes
\begin{eqnarray}
| \Psi_{d_1} \rangle &=& | N_d I_d M_d, n_d^*\kappa^*_d m^*_d ,0 \rangle\\
&\equiv& | N_d I_d M_d \rangle \otimes |n_d^*\kappa^*_d m^*_d\rangle \otimes
|0 \rangle \,, \nonumber
\end{eqnarray}
where $n_d^*$, $\kappa_d^*$ and $m^*_d$ are the quantum numbers of the 
excited electronic state. The second intermediate state following the 
fast x-ray electronic decay is characterized by the electron in the 
ground state, the excited nucleus and a photon with wavenumber $\vec{k}$ 
and polarization $\nu=1,2$,
\begin{eqnarray}
| \Psi_{d_2} \rangle &=& | N_d I_d M_d, n_d\kappa_d m_d ,\vec{k} \nu \rangle\\
&\equiv& 
| N_d I_d M_d \rangle \otimes |n_d\kappa_d m_d\rangle \otimes |\vec{k} \nu \rangle \,.
\nonumber
\end{eqnarray}
The photon is emitted in the x-ray decay of the electron from 
the excited state $| n_d^*\kappa^*_d m^*_d \rangle$ to the ground state 
$|n_d\kappa_d m_d \rangle$. The projection operator formalism presented 
in Ref.~\cite{Palffy1} is extended to account for the emission of such a photon.
The corresponding total cross section for the three-step process
$i \to d_1 \to  d_2 \to f$ can be written as
\begin{equation}
\sigma_{i \to d_1 \to d_2 \to f}(E) =\frac{2\pi^2}{p^2}\frac{A_{\rm IC}^{d_2\to
f}}{\Gamma_{d_2}}\frac{A_{\rm x-ray}^{d_1\to
d_2}}{\Gamma_{d_1}} Y^{i\to d_1}_{n}L_{d_1}(E-E_{d_1})\, ,
\label{tcs2}
\end{equation}
where $A_{\rm x-ray}$ is the electronic radiative decay rate and 
$\Gamma_{d_1}$ and $\Gamma_{d_2}$ are the widths of the two intermediate 
states. The width $\Gamma_{d_1}$ of the doubly-excited state $d_1$ is 
given as the sum of the nuclear and electronic widths, 
$\Gamma_{d_1}=\Gamma_{\gamma}+\Gamma_{\rm IC}+ \Gamma_{\rm x-ray}$, and 
can be approximated as $\Gamma_{d_1}\simeq\Gamma_{\rm x-ray}$ due to 
the difference of magnitude of the electronic and nuclear widths. The
natural width of the nuclear excited state determines the width of the second 
intermediate state $\Gamma_{d_2}$. The Lorentz profile is characterized 
in this case by the width of the first intermediate state, 
$\Gamma_{d_1}\simeq\Gamma_{\rm x-ray}$. Because of the large lifetime of 
the nuclear excited state, in Eq.~(\ref{tcs2}) an additional second 
term standing for the process in which the nuclear decay occurs prior to 
the electronic decay can be neglected.

The integration of the cross section over the continuum electron energy 
gives the resonance strength $S$ for a given capture process. In the 
case of the two-step process $i\to d\to f$ described by the total cross 
section in Eq.~(\ref{tcs1}), the continuum electron momentum $p$ and 
thus the NEEC rate $Y_n^{i \to d}$ are practically constant in the 
energy interval defined by the very narrow nuclear width. Since the 
Lorentz function is normalized to unity, the resonance strength can be 
written as
\begin{equation}
S = \frac{2\pi^2}{p^2}
\frac{A_{\rm IC}^{d \to f} Y_n^{i \to d}}{\Gamma_d} \, .
\label{rs1}
\end{equation}
For the more complicated three-step process involving NEEC into an 
excited electronic state followed by x-ray emission and IC of the 
captured electron, the resonance strength is obtained by integrating the 
total cross section given in Eq.~(\ref{tcs2}), and has the expression
\begin{equation}
S = \frac{2\pi^2}{p^2}\frac{A_{\rm IC}^{d_2\to
f}}{\Gamma_{d_2}}\frac{A_{\rm x-ray}^{d_1\to
d_2}}{\Gamma_{d_1}} Y^{i\to d_1}_{n}
 \, .
\label{rs2}
\end{equation}
Similarly, since the transition width $\Gamma_{d_1}$ is still much 
smaller than the continuum electron energy, we have assumed here 
resonance values for the momentum $p$ and the NEEC rate $Y_n^{i \to d}$.

\subsection{The electron-nucleus interaction}
%%%%%%%%%%%%%%%%%%%%%%%%%%%%%%%%%%%%%%%%%%%%%%

The NEEC rates in Eqs.~(\ref{tcs1}) and (\ref{tcs2}) are proportional to 
the squared matrix elements of the electric and magnetic 
electron-nucleus interactions and have the expression \cite{Palffy1}
\begin{eqnarray}\label{Yrate}
&&Y_n^{i \to d}=\frac{2\pi}{2(2I_i+1)}\sum_{{M_{i}} m_{s_i}}\sum_{M_{d} m_d}
\int d\Omega_p \rho_i\\
&&
\times |\langle N_d I_d M_d,n_d\kappa_d m_d,0|H_{en}+H_{magn}|N_i I_i M_i,\vec{p}_im_{s_i}^+,0\rangle|^2 \, . \nonumber
\end{eqnarray}
Here, the integral is performed over the incoming electron direction 
$\Omega_p$ and $\rho_i$ denotes the density of the electronic continuum states
($\rho_i=1$ in the units applied here). The electron-nucleus interaction Hamiltonians
$H_{en}$ and $H_{magn}$ describe the electric and magnetic transitions of the 
nucleus, respectively. We adopt the Coulomb gauge for the 
electron-nucleus interaction $H_{en}$, since it allows the separation of 
the dominant Coulomb attraction between the electronic and nuclear 
degrees of freedom,
\begin{equation}\label{eq:coulomb}
H_{en}=\int d^3r_n\frac{\rho_n(\vec{r}_n)}{|\vec{r}_e-\vec{r}_n|}\, .
\end{equation} 
Here, $\rho_n(\vec{r}_n)$ is the nuclear charge density and the 
integration is performed over the whole nuclear volume. The magnetic 
interaction Hamiltonian accounts for the recombination of the free 
electron by exchanging a virtual transverse photon. In the limit of
long exchange photon wavelength, the magnetic interaction Hamiltonian is
approximated by
\begin{equation}
H_{magn} = - \frac{1}{c} \vec{\alpha} \int d^3r_n 
\frac{\vec{j}_n(\vec{r}_n)}{|\vec{r}-\vec{r}_n|}\, ,
\label{hmagn}
\end{equation}
where  $\vec{j}_n(\vec{r}_n)$ is the nuclear current vector and $\vec{\alpha}$ is the vector
of Dirac matrices.

For describing the nucleus we use a collective model \cite{Greiner} in 
which the excitations of the nucleus are assumed to be vibrations or rotations of the 
nuclear surface. The expressions of the nuclear charge density and 
current in the two interaction Hamiltonians can then be written in terms 
of nuclear collective coordinates by means of a nuclear surface 
parameterization. Since the details of the calculation of the 
interaction Hamiltonian matrix elements are given elsewhere 
\cite{Palffy1,Palffy2}, here we only present their final expressions. 
The NEEC transition probability per unit time is given by
\begin{eqnarray}
\label{aprate}
Y^{(e)}_n&=&\frac{4\pi^2\rho_i}{(2L+1)^2} B (EL,I_i\!\to \!I_d)(2j_d+1)\\
&\times&
\sum_\kappa |R^{(e)}_{L,\kappa_d,\kappa}|^2\
%C\left(j_d\ L\ j;\frac{1}{2}\ 0\ \frac{1}{2}\right)^2\, , \nonumber
(2j+1) \left(\begin{array}{ccc} j_d&j&L\\\frac{1}{2}&-\frac{1}{2}&0\end{array}\right)^2
\, , \nonumber
\end{eqnarray}
for electric transitions of multipolarity $L$. The quantity 
\begin{equation}
B(\lambda L,I_i\!\to \! I_d)=
\frac{1}{2I_i+1}|\langle N^* I_d\|\mathcal{M}_L\|NI_i\rangle |^2
\end{equation}
represents the reduced  nuclear transition probability, where 
$\lambda$ stands for electric ($E$) or magnetic ($M$) and $\mathcal{M}$ 
is the corresponding multipole moment operator. We have denoted by 
$R^{(e)}_{L,\kappa_d,\kappa}$ the electronic matrix element
\begin{equation}
R^{(e)}_{L,\kappa_d,\kappa}=\int_0^\infty drr^{-L+1}\Big(f_{n_d\kappa_d}(r)
f_{E_c\kappa}(r)+g_{n_d\kappa_d}(r)g_{E_c\kappa}(r)\Big)\ ,
\label{radial1}
\end{equation}
where $g_{E_c\kappa}(r)$ and $f_{E_c\kappa}(r)$ are the large and small radial components of the 
relativistic continuum electron wave function in Eq.~(\ref{cont}), 
respectively, and $g_{n_d\kappa_d}(r)$ and $f_{n_d\kappa_d}(r)$ are the corresponding components 
of the bound Dirac wave functions as in Eq. (\ref{bound}). The last factor
in the summand of Eq. (\ref{aprate}) is a 3-$j$ symbol.
For magnetic 
transitions of multipolarity $L$, the NEEC rate has the expression
\begin{eqnarray}
{Y_n^{(m)}}&=&\frac{4\pi^2\rho_i}{L^2(2L+1)^2}B(ML,I_i\!\to \! I_d)(2j_d+1)
\nonumber \\
&\times&
\sum_{\kappa}
\left|R^{(m)}_{L,\kappa_d,\kappa}\right|^2
(2j+1)(\kappa_d+\kappa)
\left(\begin{array}{ccc} j_d&j&L\\\frac{1}{2}&-\frac{1}{2}&0\end{array}\right)^2 ,
\end{eqnarray}
where we introduced the following notation for the radial integral:
\begin{equation}
R^{(m)}_{L,\kappa_d,\kappa}=\int_0^{\infty}dr r^{-L+1}\bigg
(g_{n_d\kappa_d}(r)f_{E_c\kappa}(r)+\
f_{n_d\kappa_d}(r)g_{E_c\kappa}(r)\bigg) \,.
\label{radial2}
\end{equation}
The radial expressions $R^{(e)}_{L,\kappa_d,\kappa}$ and 
$R^{(m)}_{L,\kappa_d,\kappa}$ are integrated numerically.

%%%%%%%%%%%%%%%%%%%%%%%%%%%%%%%%%%%%%%%%%%%%%%%%%%%%%%%%%%%%%%%%%%%%%%%

\section{\label{results} Numerical results}

%%%%%%%%%%%%%%%%%%%%%%%%%%%%%%%%%%%%%%%%%%%%%%%%%%%%%%%%%%%%%%%%%%%%%%%

We calculate NRES total cross sections and resonance strengths for a 
number of systems involving highly charged ions of stable or long-lived 
ground state isotopes. In the first place we investigate the two-step 
process of NRES in which electron capture occurs into the electronic 
ground state and consequently the NEEC and IC bound electronic states 
coincide, as depicted in Figure~\ref{scheme1}. NEEC occurring in bare 
ions is considered for $E2$ transitions from the $0^+$ ground states to 
the first $2^+$ excited nuclear states of $^{154}_{64}\mathrm{Gd}$, 
$^{164}_{66}\mathrm{Dy}$, $^{170}_{68}\mathrm{Er}$, 
$^{174}_{70}\mathrm{Yb}$, $^{178}_{72}\mathrm{Hf}$ and 
$^{180}_{74}\mathrm{W}$. The energies of the excited nuclear levels 
$E_{\rm exc}$ as well as the reduced transition probabilities $B(E2)$, 
needed for the calculation of the natural width of the nuclear excited 
state and the NEEC  rate, are taken from 
Ref.~\cite{Raman}. The $M1$ transitions between the ground states and the 
first excited states of $^{155}_{64}\mathrm{Gd}$, 
$^{165}_{67}\mathrm{Ho}$, $^{173}_{70}\mathrm{Yb}$, 
$^{175}_{71}\mathrm{Lu}$, $^{179}_{72}\mathrm{Hf}$, 
$^{185}_{75}\mathrm{Re}$ and $^{187}_{75}\mathrm{Re}$ are also 
considered.  For these cases, the dominant $M1$ multipolarity is 
accompanied by a weaker $E2$ component. For the NEEC and IC rates, we 
take into account the multipole mixing by considering both Hamiltonians 
$H_{en}$ corresponding to the $E2$ transition and $H_{magn}$  
 corresponding to the $M1$ transition in Eq.~(\ref{Yrate}). Due to the 
specific parity of the electronic wavefunction components, the mixed 
terms in Eq.~(\ref{Yrate}) vanish and the NEEC rate can be written as a 
sum of the partial NEEC rates for the separate $M1$ and $E2$ 
multipolarities. A similar result is obtained for the $\gamma$ decay 
rate of a mixed multipole transition. The nuclear data for the magnetic 
transitions were taken from 
\cite{gd155,ho165,yb173,lu175,hf179,re185,re187}.

In the following we compare NRES, i.e., NEEC followed by IC, with the 
case when the excited nuclear state occurring in NEEC decays 
radiatively. Resonance strengths for both scenarios involving electronic 
capture into the $K$ shell of a bare ion are compared in Table~\ref{K}. 
The resonance strengths are indexed according to the nuclear decay 
channel as $S_{\rm IC}$ and $S_{\gamma}$. In the case of NEEC followed 
by the $\gamma$ decay of the nucleus, the resonance strength is given by 
\cite{Palffy1}
\begin{equation}
S_{\gamma} = \frac{2\pi^2}{p^2}
\frac{A_{\gamma}^{d \to f} Y_n^{i \to d}}{\Gamma_d} \, ,
\end{equation}
where $A_{\gamma}^{d \to f}$ is the nuclear $\gamma$ decay rate, related 
to the reduced transition probability $B$ and nuclear excitation energy 
$E_{\rm exc}$ by \cite{Ring}
\begin{equation}
A_{\gamma}^{d \to f}=\frac{8\pi(L+1)}{L[(2L+1)!!]^2}\left(E_{\rm exc}\right)^{2L+1}B(\lambda L,I_e\!\to \!I_g)\, .
\end{equation}

For the calculation of the NEEC and IC rates, the numerical evaluation 
of the radial integrals $R_{L,\kappa_d,\kappa}$ [see 
Eqs.~(\ref{radial1}) and (\ref{radial2})] is needed. We consider 
Coulomb-Dirac wave functions for the continuum electron and wave 
functions calculated with the GRASP92 package~\cite{Par96} assuming the 
potential of a homogeneously charged nucleus for the bound electron.
The value of $R_{L,\kappa_d,\kappa}$ is not affected by finite nuclear 
size effects on the accuracy level of our calculations. Nevertheless, 
the finite size of the nucleus has a sensitive effect on the energy 
levels of the bound electron. The energy of the bound electronic state 
is calculated with GRASP92 and includes one-loop one-electron quantum 
electrodynamic (QED) terms, and in the case of many-electron bound states 
approximate QED screening corrections.

\begin{table*}[htb]
\caption{\label{K} Resonance strength comparison between NRES ($S_{\rm 
IC}$) and NEEC followed by $\gamma$ decay ($S_{\gamma}$) for various 
heavy ion collision systems involving capture of the free electron in 
the $1s_{1/2}$ orbital of bare ions. The nuclear excitation energy 
$E_{\rm exc}$, the continuum electron energy at resonance in the 
center-of-mass frame $E_c$, and the multipolarity of the transition $L$ 
are given in the second, third and fourth column, respectively. }
\begin{ruledtabular}
\begin{tabular}{lrrcrr}
Isotope & $E_{\rm exc}$(keV) & $E_{c}$(keV)   & $L$ &  $S_{\rm IC}$ (b eV)&  $S_{\gamma}$ (b eV)\\ 

\hline
$^{154}_{64}\mathrm{Gd}$ & 123.071 &64.005 &$E2$ & 1.21$\times 10^{-2}$ & 2.87$\times 10^{-2}$\\
$^{164}_{66}\mathrm{Dy}$ & 73.392 &10.318   &$E2$ & 4.93$\times 10^{-2}$ & 3.86$\times 10^{-2}$\\

$^{170}_{68}\mathrm{Er}$ & 78.591 &11.350   &$E2$ & 4.90$\times 10^{-2}$ & 4.69$\times 10^{-2}$\\

$^{174}_{70}\mathrm{Yb}$ & 76.471 &4.897   &$E2$ & 3.39$\times 10^{-3}$ & 3.61$\times 10^{-3}$\\
$^{178}_{72}\mathrm{Hf}$ & 93.180 &17.103  &$E2$ & 3.11$\times 10^{-2}$ & 4.64$\times 10^{-2}$\\

$^{180}_{74}\mathrm{W}$ & 103.557 &22.776   &$E2$ & 2.30$\times 10^{-2}$ & 4.41$\times 10^{-2}$\\

\hline
$^{155}_{64}\mathrm{Gd}$ & 60.008 &0.942   &$M1+E2$ & 8.48 & 2.19 \\
$^{165}_{67}\mathrm{Ho}$ &94.700 & 29.563   &$M1+E2$ & 1.19 & 0.88\\

$^{173}_{70}\mathrm{Yb}$ & 78.647 & 7.073   &$M1+E2$ & 3.85 & 1.31\\
$^{175}_{71}\mathrm{Lu}$ & 113.804 &40.002   &$M1+E2$ & 0.153 & 0.151\\
$^{179}_{72}\mathrm{Hf}$ & 122.7909 &46.714   &$M1+E2$ & 0.327 & 0.348\\
$^{185}_{75}\mathrm{Re}$ & 125.358 & 42.198   &$M1+E2$ & 1.74 & 1.47\\
$^{187}_{75}\mathrm{Re}$ & 134.243 &51.083    &$M1+E2$ & 1.15 & 1.18\\
\end{tabular}

\end{ruledtabular}
\end{table*}

The comparison in Table~\ref{K} shows that the resonance strengths for 
NRES and for NEEC followed by  $\gamma$ emission are typically on the 
same order of magnitude. For $E2$ transitions, the $\gamma$ decay of the 
excited nuclear state tends to dominate over the IC decay, so that 
$S_{\rm IC} <S_{\gamma}$. For magnetic dipole transitions, typically the 
NRES resonance strength values are larger than the ones for NEEC 
followed by $\gamma$ emission, culminating with the case of 
$^{155}_{64}\mathrm{Gd}$, for which $S_{\rm IC}$=8.48~b~eV and 
$S_{\gamma}$=2.19~b~eV. The difference between $S_{\rm IC}$ and 
$S_{\gamma}$ is given by the decay channel only, namely, by the decay 
rates of the nuclear excited state $A_{\rm IC}$ and $A_{\gamma}$, 
respectively. The ratio $\alpha= A_{\rm IC}/A_{\gamma}$ denotes the IC 
coefficient, whose values are calculated for a particular bound shell or 
orbital. The behaviour of $\alpha$ with respect to the capture orbital 
depends on the multipolarity of the nuclear transition. While for $E2$ 
transitions $p$ orbitals have larger $\alpha$ values, for $M1$ 
transitions the $s$ orbitals have a stronger contribution to the total 
IC coefficient. It is therefore not surprising that in Table~\ref{K}, 
with resonance strengths considering NEEC into the $1s_{1/2}$ orbital of 
bare ions, for $M1$ transitions the NRES resonance strengths  are typically 
larger than the ones for NEEC followed by $\gamma$ decay, $S_{\rm 
IC}\gtrsim S_{\gamma}$. In addition, the NRES resonance strengths for $M1$ transitions are 
substantially larger than the ones for $E2$ transitions, also due to the broader
natural line widths of the former.

For heavier even-even nuclei such as the actinides 
$^{232}_{90}\mathrm{Th}$, $^{236}_{92}\mathrm{U}$, 
$^{238}_{92}\mathrm{U}$, and $^{248}_{96}\mathrm{Cm}$, the capture into 
the $K$ shell is not possible since the binding energy of the $1s_{1/2}$ 
electron is larger than the nuclear excitation energy. These nuclei 
present first-excited $2^+$ states lying at about 40~keV above the $0^+$ 
ground state. Due to their low transition energies and large 
corresponding IC coefficients, the actinide nuclei are prospective 
candidates for NRES with recombination into the $L$ shell. In Table~\ref{U_LL}, we consider NEEC with capture 
into the $2s_{1/2}$, $2p_{1/2}$ and $2p_{3/2}$ orbitals of the ground 
state electronic configuration of the He-like ($1s_{1/2}^2$), Be-like 
($1s_{1/2}^22s^2_{1/2}$) and C-like ($1s_{1/2}^22s_{1/2}^22p_{1/2}^2$) ions, 
respectively.  NRES resonance strengths are compared to the ones of NEEC 
followed by $\gamma$ decay of the nucleus. Unlike the cases of NEEC 
occurring in bare ions, in this case the width of the nuclear excited 
state may also contain terms corresponding to the IC decay of the bound 
electrons in the initial electronic configuration. The presence of the 
$K$-shell electrons does not play any role in the nuclear decay, since 
the low energy of the nuclear transition does not allow their IC. For 
Be-like and C-like ions, however, the IC decay rates of the $2s_{1/2}$ 
and $2p_{1/2}$ orbital electrons contribute to the total width of the 
nuclear excited state. For the calculation of the 
radial wave functions for the continuum electron, we assume a total 
screening of the nuclear charge, i.e., we use Coulomb-Dirac functions 
with an effective nuclear charge $Z_{\rm eff} = Z-N$, where $N$ stands 
for the number of bound electrons.  For the bound electron wave 
functions, the electron-electron interaction is accounted for in the 
Dirac-Fock approximation.

\begin{table*}[htb]
\caption{\label{U_LL} Resonance strength comparison between NRES 
($S_{\rm IC}$) and NEEC followed by $\gamma$ decay ($S_{\gamma}$) for 
several heavy ion collision systems involving capture of the free 
electron in the $2s_{1/2}$ orbital of He-like ions, the $2p_{1/2}$ 
orbital of Be-like ions and the $2p_{3/2}$ orbital of C-like ions. The 
nuclear excitation energy $E_{\rm exc}$, the continuum electron energy 
at resonance in the center-of-mass frame $E_c$, and the capture orbital 
$nl_j$ are given in the second, third and fourth column, respectively.}
\begin{ruledtabular}
\begin{tabular}{lrrcrr}
Isotope & $E_{\rm exc}$(keV) & $E_{c}$(keV)  & $nl_j$  &  $S_{\rm IC}$ (b eV)&  $S_{\gamma}$ (b eV)\\ 
\hline
 &  &18.244  & $2s_{1/2}$ & 0.011 &$5.44\times 10^{-3}$ \\
$^{232}_{90}\mathrm{Th}$ & 49.369 &19.400 & $2p_{1/2}$ &  0.416 & 6.93$\times 10^{-3}$\\
 &  &24.010 & $2p_{3/2}$ & 0.055  & 1.95$\times 10^{-3}$\\
\hline
			&	 &12.405  & $2s_{1/2}$ & 0.033 & $7.99\times 10^{-3}$\\
$^{236}_{92}\mathrm{U}$ & 45.242 & 13.596 & $2p_{1/2}$ & 0.906 &  8.32$\times 10^{-3}$\\
 			 & 	 & 19.655 & $2p_{3/2}$ & 0.098 & 1.97$\times 10^{-3}$ \\
\hline
  			&   & 12.073 & $2s_{1/2}$ & 0.039& $9.06\times 10^{-3}$\\
$^{238}_{92}\mathrm{U}$ & 44.916 & 13.262 & $2p_{1/2}$ &  1.055 & 9.35$\times 10^{-3}$\\
  			&   &18.323  & $2p_{3/2}$ & 0.120  &2.32$\times 10^{-3}$ \\
\hline
			 &   &6.888  & $2s_{1/2}$ & 0.147 & 1.79$\times 10^{-2}$ \\
$^{248}_{96}\mathrm{Cm}$ & 43.380 &8.190 & $2p_{1/2}$ &  2.936 &  1.55$\times 10^{-2}$ \\
			 &   &14.203 & $2p_{3/2}$ & 0.240  & 2.94$\times 10^{-3}$  \\
\end{tabular}

\end{ruledtabular}
\end{table*}

We find that NRES resonance strengths for electron capture and 
scattering on the $2p$ orbitals are up to two orders of magnitude larger 
than the corresponding values for NEEC followed by $\gamma$ decay 
of the nucleus for the highly charged actinides presented in 
Table~\ref{U_LL}. The $2p$ orbitals of $^{232}_{90}\mathrm{Th}$, 
$^{236}_{92}\mathrm{U}$, $^{238}_{92}\mathrm{U}$ and 
$^{248}_{96}\mathrm{Cm}$ have a major role in the IC decay of the 
excited nuclear state. The IC coefficients corresponding to the neutral 
atom have values between $\alpha=327$ for the transition of 
$^{232}_{90}\mathrm{Th}$ and $\alpha=984$ for the one of 
$^{248}_{96}\mathrm{Cm}$. In few-electron configurations, however, the 
strongly-bound inner-shell electrons have a more pronounced influence on 
nuclear coupling to the atomic shells \cite{Philips}. One electron in 
the $2p_{1/2}$ orbital of the highly charged ion of 
$^{248}_{96}\mathrm{Cm}$, for instance, accounts already for a partial 
IC coefficient of $\alpha=188$. Thus the two orders of magnitude 
difference between $S_{\rm IC}$ and $S_{\gamma}$ in Table~\ref{U_LL} can 
be traced back to the behaviour of the IC and radiative decay rates and 
the IC coefficient. The relatively small energy of the nuclear 
transitions leads to a low radiative decay rate. On the other hand, the 
heavy actinides presented in Table~\ref{U_LL} are high-$Z$ nuclei with 
large radial electronic integrals $R_{L,\kappa_d,\kappa}$ which lead to 
significant IC rates. The largest NRES resonance strength is associated 
with capture into the $2p_{1/2}$ orbital of $^{248}_{96}\mathrm{Cm}$, 
with $S_{\rm IC}=2.94$~b~eV, which is on the same order of magnitude 
with the corresponding values presented in Table~\ref{K} for $M1$ 
nuclear transitions.

In the scenario considered so far, only few electrons fulfill the 
resonance condition due to the very narrow natural width of the nuclear 
excited state. A possibility to relax the resonance condition is given 
by the capture of the electron into an excited bound state. We consider 
therefore the more complicated three-step NRESX process depicted in 
Figure~\ref{scheme2}, in which NEEC into an excited electronic state is 
followed by $K\alpha$ x-ray emission and only subsequently by IC of 
the captured electron. Since the capture and IC electronic states do not 
coincide, the scattered free electron will have a different kinetic 
energy than the incident electronic beam, with the difference in energy 
being carried away by the x-ray photon. Furthermore, for the very heavy 
actinide nuclei where the electron capture into the $K$ shell with the 
excitation of the first collective excited level is not possible, NRESX 
among $L$ subshells turns out to have several advantages for 
experimental observation. The resonance strength for NRESX is calculated 
using the expression in Eq.~(\ref{rs2}), where the required electronic 
widths and x-ray transition rates are provided by the OSCL92 module of 
the GRASP92 package.

For the first group of isotopes in Table~\ref{K}, where capture into 
and, consequently, IC from the $K$ shell is possible, we envisage NEEC 
into the $L$ shell of bare ions. The x-ray decay of the captured 
electron to the $K$ shell occurs orders of magnitude faster than the 
nuclear de-excitation. In this case, the IC decay of the nucleus will 
follow the x-ray emission and will ionize the bound electron from the 
$K$ shell. In Tables \ref{L_E2} and \ref{L_M1} we present continuum 
electron energies $E_c$, NEEC rates $Y_n$ and NRESX resonance strengths 
$S$ for the capture into the $2s_{1/2}$, $2p_{1/2}$ and $2p_{3/2}$ 
orbitals of bare ions. Compared to the resonance strengths for the 
two-step process of NRES presented in Table~\ref{K}, the values for 
NRESX are several orders of magnitude smaller. The main reason for this 
behaviour is the dependence of the resonance strength on the momentum of 
the incoming continuum electron $p$, presented in Eq.~(\ref{tcs2}). All 
isotopes in Tables \ref{L_E2} and \ref{L_M1} have nuclear excitation 
energies that allow NEEC into the $K$ shell. Since the capture occurs 
however into the $L$ shell, the continuum electron energy at  the
resonance  has large values, starting from the energy difference 
between the $L$ and $K$ shells. As shown in the third column of Tables 
\ref{L_E2} and \ref{L_M1}, the continuum electron energy has values 
between approximately 45~keV for the case of $^{155}_{64}\mathrm{Gd}$ 
and going up to 114~keV for $^{187}_{75}\mathrm{Re}$.

\begin{table*}[htb]
\caption{\label{L_E2}
Resonance strengths $S$ for NRESX: electron 
recombination into the $L$ shell orbitals of bare ions followed by $K$-shell IC. 
The nuclear transition multipolarity is $E2$. $E_{\rm exc}$ denotes the 
nuclear exitation energy, $E_c$ is the continuum electron energy in 
the center-of-mass frame and the capture orbital is denoted by $nl_j$. 
In the fifth column we present the NEEC rate $Y_{n}$ and in the seventh the maximum
value of the convoluted cross section, $\tilde{\sigma}_{\rm max}$. The last column
contains total cross sections $\sigma_{bs}$ for the competing process of
bremsstrahlung.
}
\begin{ruledtabular}
\begin{tabular}{lrrclrrc}
Isotope & $E_{\rm exc}$(keV) & $E_{c}$(keV)  & $nl_j$ &   $Y_{n}$ (1/s) & 	$S$ (b eV) 	& $\tilde{\sigma}_{\rm max}$ (b) & $\sigma_{bs}$ (b) \\ 
\hline
			 &         & 108.077 & $2s_{1/2}$   & $6.86\cdot 10^7$ & $8.44\times 10^{-4}$ & $3.36\times 10^{-5}$&  \\
$^{154}_{64}\mathrm{Gd}$ & 123.071 & 108.063 & $2p_{1/2}$   & $1.14\cdot 10^8$ & $1.40\times 10^{-3}$ & $1.24\times 10^{-4}$& 12.7 \\
			 &         & 108.946 & $2p_{3/2}$   & $1.51\cdot 10^8$ & $1.84\times 10^{-3}$ & $1.82\times 10^{-4}$&  \\
\hline                                                                                                  
			 &         & 57.365  & $2s_{1/2}$   & $2.36\cdot 10^7$ & $1.08\times 10^{-3}$ & $4.30\times 10^{-5}$&  \\
$^{164}_{66}\mathrm{Dy}$ & 73.392  & 57.348 & $2p_{1/2}$    & $1.85\cdot 10^8$ & $8.49\times 10^{-3}$ & $6.65\times 10^{-4}$& 13.5 \\
			 &         & 58.357 & $2p_{3/2}$    & $2.54\cdot 10^8$ & $1.14\times 10^{-2}$ & $1.01\times 10^{-3}$&  \\
\hline                                                                                                  
			 &         & 61.484  & $2s_{1/2}$   & $2.87\cdot 10^7$ & $1.11\times 10^{-3}$ & $4.42\times 10^{-5}$&  \\
$^{170}_{68}\mathrm{Er}$ & 78.591  & 61.468 & $2p_{1/2}$    & $2.33\cdot 10^8$ & $9.04\times 10^{-3}$ & $6.26\times 10^{-4}$& 14.3 \\
			 &         & 62.616 & $2p_{3/2}$    & $3.10\cdot 10^8$ & $1.18\times 10^{-2}$ & $9.30\times 10^{-4}$&  \\
\hline                                                                                                  
			 &         & 58.241  & $2s_{1/2}$    & $9.88\cdot 10^5$ & $3.85\times 10^{-5}$ &$1.53\times 10^{-6}$ &  \\
$^{174}_{70}\mathrm{Yb}$ & 76.471  & 58.222  & $2p_{1/2}$    & $1.10\cdot 10^7$ & $4.30\times 10^{-4}$ &$2.63\times 10^{-5}$ & 15.2 \\
			 &         & 59.526  & $2p_{3/2}$    & $1.44\cdot 10^7$ & $5.48\times 10^{-4}$ &$3.91\times 10^{-5}$ &  \\
                                                                                                        
\hline                                                                                                  
			 &         & 73.780  & $2s_{1/2}$    & $3.60\cdot 10^7$ & $9.06\times 10^{-4}$ &$3.61\times 10^{-5}$ &  \\
$^{178}_{72}\mathrm{Hf}$ &  93.180 & 73.759  & $2p_{1/2}$    & $2.76\cdot 10^8$ & $6.94\times 10^{-3}$ &$3.79\times 10^{-4}$ & 16.0 \\
			 &         & 75.235  & $2p_{3/2}$    & $3.44\cdot 10^8$ & $8.47\times 10^{-3}$ &$5.43\times 10^{-4}$ &  \\
                                                                                                        
\hline                                                                                                  
			 &         & 82.939  & $2s_{1/2}$    & $4.22\cdot 10^7$ & $8.00\times 10^{-4}$ &$3.19\times 10^{-5}$ &  \\
$^{180}_{74}\mathrm{W}$  & 103.557 &  82.915 & $2p_{1/2}$    & $2.95\cdot 10^8$ & $5.59\times 10^{-3}$ &$2.72\times 10^{-4}$ & 16.9 \\
			 &         &  84.582 & $2p_{3/2}$    & $3.54\cdot 10^8$ & $6.56\times 10^{-3}$ &$3.85\times 10^{-4}$ &  \\
\end{tabular}
\end{ruledtabular}
\end{table*}

\begin{table*}[htb]
\caption{\label{L_M1}
Same as Table \ref{L_E2} for isotopes with $M1$ nuclear transitions.
}
\begin{ruledtabular}
\begin{tabular}{lrrcllcc}
Isotope & $E_{\rm exc}$(keV) & $E_{c}$(keV)  & $nl_j$ &  $Y_{n}(1/s)$ & $S$ (b eV) & $\tilde{\sigma}_{\rm max}$ (b) & $\sigma_{bs}$ (b)\\ 
\hline
			 &         & 45.014 &  $2s_{1/2}$  &$3.12\cdot 10^8$ &  $2.61\times 10^{-2}$ &$1.04\times 10^{-3}$ &  \\
$^{155}_{64}\mathrm{Gd}$ & 60.008  & 45.001 &  $2p_{1/2}$  &$8.37\cdot 10^7$ &  $7.01\times 10^{-3}$ &$6.22\times 10^{-4}$ & 12.7 \\
			 &         & 45.883 & $2p_{3/2}$   &$8.39\cdot 10^7$ &  $6.89\times 10^{-3}$ &$6.82\times 10^{-4}$ &  \\
                                                                                                      
\hline                                                                                                
			 &         & 78.138  &  $2s_{1/2}$  &$2.00\cdot 10^9$ & $6.75\times 10^{-2}$ &$2.69\times 10^{-3}$ &  \\
$^{165}_{67}\mathrm{Ho}$ &  94.700 & 78.122  &  $2p_{1/2}$  &$2.71\cdot 10^8$ & $9.17\times 10^{-3}$ &$6.73\times 10^{-4}$ & 13.9 \\
			 &         & 79.198  & $2p_{3/2}$   &$1.47\cdot 10^8$ & $4.89\times 10^{-3}$ &$4.08\times 10^{-4}$ & \\
                                                                                                      
\hline                                                                                                
			 &	   &  60.417 &  $2s_{1/2}$  &$1.18\cdot 10^9$ & $6.85\times 10^{-2}$ &$2.73\times 10^{-3}$ & \\
$^{173}_{70}\mathrm{Yb}$ &  78.647 &  60.398 &  $2p_{1/2}$  &$2.67\cdot 10^8$ & $1.54\times 10^{-2}$ &$9.40\times 10^{-4}$ & 15.2 \\
 			 &	   &  61.702 &  $2p_{3/2}$  &$2.17\cdot 10^8$ & $1.22\times 10^{-2}$ &$8.70\times 10^{-4}$ & \\

\hline                                                                                                
			 &	   & 94.995  &  $2s_{1/2}$  &$4.04\cdot 10^8$ & $9.72\times 10^{-3}$ &$3.87\times 10^{-4}$ & \\
$^{175}_{71}\mathrm{Lu}$ & 113.804 & 94.975  &  $2p_{1/2}$  &$1.39\cdot 10^8$ & $3.35\times 10^{-3}$ &$1.94\times 10^{-4}$ & 15.6 \\
 			 &	   &  96.363 &  $2p_{3/2}$  &$1.32\cdot 10^8$ & $3.13\times 10^{-3}$ &$2.12\times 10^{-4}$ & \\
                                                                                                      
\hline                                                                                                
			 &	   & 103.391 &  $2s_{1/2}$  &$1.07\cdot 10^9$ & $2.26\times 10^{-2}$ &$9.00\times 10^{-4}$ & \\
$^{179}_{72}\mathrm{Hf}$ & 122.791 & 103.370 &  $2p_{1/2}$  &$2.06\cdot 10^8$ & $4.34\times 10^{-3}$ &$2.37\times 10^{-4}$ & 16.0 \\
 			 &	   & 104.846 &  $2p_{3/2}$  &$1.39\cdot 10^8$ & $2.90\times 10^{-3}$ &$1.86\times 10^{-4}$ & \\
                                                                                                      
\hline                                                                                                
			 &         & 104.112 & $2s_{1/2}$  &$4.74\cdot 10^{9}$& $1.11\times 10^{-1}$ &$4.42\times 10^{-3}$ & \\
$^{185}_{75}\mathrm{Re}$ &  125.35 & 104.086 &  $2p_{1/2}$ &$6.15\cdot 10^8$  & $1.43\times 10^{-2}$ &$6.60\times 10^{-4}$ & 17.4 \\
			 &         & 105.857 & $2p_{3/2}$  &$2.22\cdot 10^8$  & $5.11\times 10^{-3}$ &$2.84\times 10^{-4}$ &  \\
\hline                                                                                                
                                                                                                      
			 &         & 112.996 & $2s_{1/2}$  &$4.20\cdot 10^{9}$& $8.17\times 10^{-2}$ &$3.25\times 10^{-3}$ & \\
$^{187}_{75}\mathrm{Re}$ & 134.24  & 112.970 & $2p_{1/2}$  &$5.21\cdot 10^8$  & $1.01\times 10^{-2}$ &$4.60\times 10^{-4}$ & 17.4 \\
			 &         & 114.741 & $2p_{3/2}$  &$1.68\cdot 10^8$  & $3.22\times 10^{-3}$ &$1.79\times 10^{-4}$ & \\

\end{tabular}
\end{ruledtabular}
\end{table*}

\begin{table*}[htb]
\caption{\label{U_ML} Resonance strengths $S$ for NRESX with capture 
into the $L$- and $M$-shell $p$ orbitals of He-like ions followed by the 
intra-shell or $L \alpha$ radiative decay to the $2s$ state and IC of 
the bound electron. The nuclear transition multipolarity is $E2$. 
$E_{\rm exc}$ denotes the nuclear exitation energy, $E_{c}$ is the 
continuum electron energy in the center-of-mass frame and $nl_j$ stands 
for the capture orbital.
%The last column contains bremsstrahlung total
%cross sections $\sigma_{bs}$.
}
\begin{ruledtabular}
\begin{tabular}{lrcccccl}
Isotope & $E_{\rm exc}$(keV) & $nl_j$ & $E_{c}$(keV)  &$S$ (b eV)  &  $nl_j$ & $E_{c}$(keV) &$S$ (b eV) \\
\hline
$^{232}_{90}\mathrm{Th}$  &49.369 &  $2p_{1/2}$ & 18.517 &  0.337 & $3p_{1/2}$   & 36.101  &0.059  \\

$^{232}_{90}\mathrm{Th}$  & 49.369 &   $2p_{3/2}$& 22.270 & 0.291 & $3p_{3/2}$ &  37.223 & 0.064  \\
\hline
$^{236}_{92}\mathrm{U}$  & 45.242 &   $2p_{1/2}$& 12.688 & 0.894 & $3p_{1/2}$  & 31.270  & 0.125  \\

$^{236}_{92}\mathrm{U}$  & 45.242 &   $2p_{3/2}$& 16.872 & 0.683 & $3p_{3/2}$ &  32.519  &  0.131  \\
\hline
$^{238}_{92}\mathrm{U}$  & 44.916 &   $2p_{1/2}$ & 12.362 & 1.049 & $3p_{1/2}$ & 30.941  & 0.144  \\

$^{238}_{92}\mathrm{U}$  & 44.916 &   $2p_{3/2}$&  16.540 & 0.797 & $3p_{3/2}$ & 32.190 &  0.151   \\
\hline
$^{248}_{96}\mathrm{Cm}$  & 43.380 &   $2p_{1/2}$& 7.210  & 3.411 & $3p_{1/2}$ & 27.927 & 0.306  \\

$^{248}_{96}\mathrm{Cm}$  & 43.380 &   $2p_{3/2}$& 12.376 & 1.910 & $3p_{3/2}$ & 29.469  & 0.299  \\

\end{tabular}

\end{ruledtabular}
\end{table*}

The case of the heavy actinide nuclei with low-lying first excited 
states also offers other scenarios for NRESX. Since IC of the $K$-shell 
electrons is energetically forbidden, we consider in the following 
capture into initially He-like ions. A first scenario considers NEEC 
into the $2p$ orbitals of He-like ions, followed by the fast x-ray decay 
of the captured electron to the $2s_{1/2}$ state. The x-ray transition 
for the considered highly charged ions occurs several orders of 
magnitude faster than the nuclear decay. IC will therefore follow the 
electronic transition and ionize the $2s_{1/2}$ electron. The initial 
and final continuum electron energies are then given by the 
corresponding energies of the IC and capture $L$ subshells.  In a 
similar manner, one can envisage the NEEC into the $3p$ orbitals of 
He-like ions, with the subsequent decay of the captured electron to the 
$2s$ ground state. In Table~\ref{U_ML} we present continuum electron 
energies and NRESX resonance strengths for NEEC occurring into the $2p$ 
and $3p$ orbitals of He-like ${\rm Th}^{88+}$, ${\rm U}^{90+}$, and 
${\rm Cm}^{94+}$ ions. The resonance strengths for NEEC into the $2p$ 
orbitals are larger than the ones for capture into the $3p$ orbitals, 
due to the smaller electron momentum values and different overlap of 
the electronic wave functions with the nuclear matter. The largest 
resonance strength value is the one for NRESX with capture into the 
$2p_{1/2}$ orbital of the initially He-like ${\rm Cm}^{94+}$ ion, 
namely, $S=3.41$~b eV.

The initial and final states of NRESX coincide with those of
bremsstrahlung, where a photon is directly emitted in a continuum-continuum 
transition. Bremsstrahlung is therefore a background process
which may complicate the observation of NRESX. In Tables \ref{L_E2} and \ref{L_M1}
we give total radiation cross section values calculated within a nonrelativistic
approximation (photon energy $\ll$ electron rest energy)~\cite{Koch&Motz} for 
orientation. For comparison,
the maximum values of the NRESX cross sections convoluted with a 10~eV width Gaussian
electron energy distribution  are given ($\tilde{\sigma}_{\rm max}$).
Bremsstrahlung cross sections are typically 4-6 orders of magnitude
larger than the NRESX cross sections at resonance.
However, since the lifetime of the NRESX process is dominated by long nuclear
mean-lives, it occurs on a much longer time scale than bremsstrahlung.
This fact may be exploited in a possible observation of nuclear-resonant electron scattering.

%%%%%%%%%%%%%%%%%%%%%%%%%%%%%%%%%%%%%%%%%%%%%%%%%%%%%%%%%%
% about experiment
%%%%%%%%%%%%%%%%%%%%%%%%%%%%%%%%%%%%%%%%%%%%%%%%%%%%%%%%%%

Regarding the possible experimental observation of NRES, the high 
sensibility of electron spectrometry to strong magnetic fields restricts 
the choice of the experimental setup. The presence of magnetic fields 
 in electron coolers of 
storage rings perturbs the electron trajectory and makes the detection 
of scattered electrons very difficult. At the storage ring facility at 
the GSI, electron scattering experiments have been 
performed using a gas target as electron target 
\cite{Hagmann_spectrometer,NIMB,Hagmann_prl}. The accelerated ions are cycling in 
the Experimental Storage Ring (ESR) with velocities close to the speed of 
light and are passing through a gas target with electron densities of 
$10^{12}-10^{14}$ electrons per cm$^2$. The quasifree electrons are then 
scattered by the highly charged ions. The main drawback related to the 
use of gas targets is the  nuclear Coulomb excitation that occurs due
to the target nuclei.

For an envisaged NRES experiment, when the resonance energy condition 
is fulfilled, the quasifree electrons can be captured by the fast ion 
with the simultaneous excitation of the nucleus, being then carried away 
from the gas target. After a time interval corresponding to the mean 
lifetime of the nuclear excited state in the actual electronic configuration
of the highly charged ion, the electron is expelled by the 
ion and can be detected using an electron spectrometer 
\cite{Hagmann_spectrometer}. The energy of the electron will be given by 
the transformation of the final continuum electron energy $E_c$ from the 
center-of-mass frame to the laboratory frame \cite{Eichler}, 
\begin{equation}
E_c^{lab}/c=\gamma(E_c^{cm}/c+\beta 
p^{cm}\mathrm{cos}\theta) \, .
\end{equation}
The emitted electron is 
characterized in the center-of-mass system by the momentum $p^{cm}$ with 
a direction determined by the polar angle $\theta$ with respect to the 
$z$ axis. In the equation above, $c$ stands for the speed of light and 
$\beta$ and $\gamma$ are the reduced velocity and the Lorentz factor of 
the ion, respectively. The electron spectrometer actually detects the 
electrons emitted in the forward direction \cite{Hagmann_prl,Hagmann_spectrometer}.

The initial and final states of the scattering process in this scenario 
are the same as the ones of the non-resonant process of electron capture 
to the continuum (ECC)~\cite{ecc1,ecc2,ecc_doris}. This process is one 
of the major sources of background for NRES, and due to the identical 
initial and final states, quantum interference between the two processes 
may occur. However, an investigation of the corresponding time scales 
for the two processes reveals that in contrast to the nuclear 
lifetime-dependent NRES, ECC occurs much faster. Time-discrimination 
spectroscopy, as it has been proposed in Ref.~\cite{lifetime_prolongation} 
for NEEC followed by $\gamma$ decay of the nucleus, can therefore reduce 
substantially the ECC background. Similarly to the concept presented in Ref.
\cite{lifetime_prolongation}, the different time scales of NRES and ECC 
have as a result a spatial separation of the electron emissions in a 
storage ring experiment. While the ECC photons will be emitted almost
instantaneously in the region of the gas target, internal conversion 
will only occur later, after the ions have already travelled a certain 
distance in the ring. For the present electron spectrometer, where
all forward-emitted electrons are detected approximately 90~cm after the gas target 
\cite{Hagmann_prl,Hagmann_spectrometer}, the separation of the signal and 
background events  is challenging and requires a special extension of the experimental setup.

Particularly interesting is the case of NRESX occcuring into the $2p$ 
orbitals of He-like ions of heavy actinides, with resonance strengths 
presented in Table~\ref{U_ML}. The captured electron undergoes a fast 
x-ray decay to the $2s_{1/2}$ orbital (the decay rates 
%
%calculated with the OSCL92 module of the GRASP92 package 
%
are $1.95\times 10^{10}$ 
s$^{-1}$ for the $2p_{1/2} \to 2s_{1/2}$ transition and $8.86\times 
10^{14}$ s$^{-1}$ for the $2p_{3/2} \to 2s_{1/2}$ transition). The IC rate for the 
$2s$ orbital electron is much smaller than the one for $2p$ orbital 
electrons, so that the nuclear lifetime in the case of the $1s^2 2s$
configuration is longer that the one for the $1s^2 2p$ capture
configurations. The nuclear mean-lives of the $1s^2 2s$ 
ion configuration of the four studied heavy actinides have values 
between $\tau$=13~ns for $^{248}_{96}\mathrm{Cm}$ and $\tau$=50~ns for 
$^{232}_{90}\mathrm{Th}$, corresponding to a spatial separation of 
approximately 13 to 50~cm.

%%%%%%%%%%%%%%%%%%%%%%%%%%%%%%%%%%%%%%%%%%%%%%%%%%%%%%%%%%%%%%%%%%%%%%%%%%

\section{\label{sum} Summary}

%%%%%%%%%%%%%%%%%%%%%%%%%%%%%%%%%%%%%%%%%%%%%%%%%%%%%%%%%%%%%%%%%%%%%%%%%%

We have considered nuclear-resonant electron scattering,
i.e. nuclear excitation by electron capture followed by internal conversion,
focusing on finding prospective isotopes 
for a possible experimental observation of the process. Theoretical total 
cross sections and resonance strengths for a number of capture scenarios 
and collision systems have been presented.

In the first place, we have investigated the process of NRES involving 
$E2$ and $M1$ nuclear transitions with electron recombination into the 
electronic ground state. A comparison with resonance strengths for NEEC 
followed by the radiative decay of the nucleus shows that the two 
processes are typically on the same order of magnitude. For the specific 
cases of the heavy actinides studied, the IC nuclear decay channel 
prevails, and the NRES resonance strengths are between one and two 
orders orders or magnitude larger.

A second scenario in which the electronic capture occurs into an excited 
electronic state and is followed by x-ray emission has also been 
investigated. Due to the large  width of the excited electronic 
state, the continuum electron resonance energy condition is 
significantly relaxed. We have found that for the heavy actinide 
isotopes, NRESX with electronic capture and IC from different subshells of the $L$ 
shell presents large resonance strength values. Furthermore, the 
possible experimental observation of NRES in storage rings, e.g. at 
the present and future ESR 
facility of the GSI Darmstadt has been discussed, devoting special attention to 
the electron target setup. A time-discrimination measurement at the 
ESR could be used for discerning the process of NRES from the background of other 
atomic physics processes, such as ECC. The most promising candidates for 
time-discrimination measurements were found to be the heavy actinide 
nuclei in a NRESX scenario involving the $L$-shell orbitals. While the 
calculated resonance strengths, on the order of 1~b\ eV, still make the 
 observation of the NRES effect challenging, the advent of 
the new storage ring facility at GSI and the reported interest for 
electron spectroscopy experiments in the relativistic regime are strong 
arguments for the need of consistent theoretical predictions and 
experimental scenarios.

%%%%%%%%%%%%%%%%%%%%%%%%%%%%%%%%%%%%%%%%%%%%%%%%%%%%%%%%%%%%%%%%%%%%%%%%%%

\begin{acknowledgments}

The authors would like to thank Alexander Voitkiv for helpful comments. 
A.P. is indebted to Stefan Schippers,  Siegbert Hagmann and Christophor Kozhuharov for 
discussions concerning the experimental issues mentioned in this paper.

\end{acknowledgments}

%%%%%%%%%%%%%%%%%%%%%%%%%%%%%%%%%%%%%%%%%%%%%%%%%%%%%%%%%%%%%%%%%%%%%%%%%%
%\appendix
%\section{\label{A}}
%%%%%%%%%%%%%%%%%%%%%%%%%%%%%%%%%%%%%%%%%%%%%%%%%%%%%%%%%%%%%%%%%%%%%%%%%%
\bibliography{res}

\end{document}